\documentclass{article}
\usepackage{spconf,amsmath,graphicx}
\usepackage[table]{xcolor}
\usepackage{multirow}
\usepackage{cite}
\usepackage{booktabs}
\usepackage{hyperref}
\usepackage{diagbox} 
\usepackage{pifont}
\usepackage{url}
\usepackage[bottom]{footmisc}
\usepackage{balance}

\title{ESDD 2026: Environmental Sound Deepfake Detection Challenge Evaluation Plan}
\name{Han Yin$^{1}$, Yang Xiao$^{2,3}$, Rohan Kumar Das$^{3}$, Jisheng Bai$^{4}$ and Ting Dang$^{2}$}

\address{$^1$School of Electrical Engineering, KAIST, Daejeon, Republic of Korea\\
$^2$University of Melbourne, Australia \\
$^3$Fortemedia Singapore, Singapore \\
$^4$Xi'an University of Posts \& Telecommunications, Xi'an, China
}

\begin{document}
\ninept
\maketitle
\begin{abstract}
Recent advances in audio generation systems have enabled the creation of highly realistic and immersive soundscapes, which are increasingly used in film and virtual reality. However, these audio generators also raise concerns about potential misuse, such as generating deceptive audio content for fake videos and spreading misleading information.
Existing datasets for environmental sound deepfake detection (ESDD) are limited in scale and audio types.
To address this gap, we have proposed EnvSDD\footnote{EnvSDD Dataset: https://envsdd.github.io/}, the first large-scale curated dataset designed for ESDD, consisting of 45.25 hours of real and 316.7 hours of fake sound.
Based on EnvSDD, we are launching the Environmental Sound Deepfake Detection Challenge\footnote{ESDD 2026: https://sites.google.com/view/esdd-challenge}\footnote{Challenge Baseline: https://github.com/apple-yinhan/EnvSDD}.
Specifically, we present two different tracks: ESDD in Unseen Generators and Black-Box Low-Resource ESDD, covering various challenges encountered in real-life scenarios.
The challenge will be held in conjunction with the 2026 IEEE International Conference on Acoustics, Speech, and Signal Processing (ICASSP 2026)\footnote{ICASSP 2026: https://2026.ieeeicassp.org/}.

\end{abstract}

\section{Introduction}
The rapid advancement of audio generation models has made it increasingly easy to create highly realistic deepfake speech, music and environmental sound, which can be exploited to spread misinformation and mislead the public \cite{deepfake-review-1,deepfake-review-2}. 
In recent years, speech and singing voice deepfake detection has made a significant milestone, leading to various state-of-the-art (SOTA) models and benchmarks \cite{zhang2024svdd,add_challenge,asvspoof2017}.
However, how to effectively detect deepfake environmental sound clips (e.g., fake gun shots and fire alarms) is underexplored, with a lack of large-scale and publicly available datasets.


To address this issue, in our previous work, EnvSDD \cite{envsdd}, we presented the first large-scale dataset for environmental sound deepfake detection (ESDD). The dataset contains sound clips sampled from different real-life scenarios and deepfake clips generated with various sound generators, including two distinct generation paradigms, text-to-audio (TTA) and audio-to-audio (ATA).
A key insight in EnvSDD is that the trained ESDD models have a significant drop in detection performance when applied to unseen generators. This trend is similar with that in speech and singing voice deepfake detection domains \cite{zang2024ctrsvdd,asvspoof_2021,asvspoof5}.
Although the detection performance on unseen generators can be improved by introducing BEATs \cite{beats}, a model pre-trained on large-scale environmental sound using self-supervised learning, the improvement remains limited, with a reduction in equal error rate ranging from 0.3\% to 2.44\%.

To promote the development of ESDD, we introduce the ESDD challenge, the first dedicated research initiative specifically designed to explore and address the unique challenges associated with detecting AI-generated deepfake sound in the environmental sound domain. Specifically, this initiative aims to foster innovation, benchmark progress, and encourage the development of robust detection methods tailored to real-world acoustic scenarios.

\section{Challenge Overview}
\subsection{Challenge Tracks}
In EnvSDD, we employed five TTA models and two ATA models to generate the fake sound clips. During training, the detection models were exposed to only a subset of these generators. For evaluation, the trained models had to generalize to unseen generators. This setup closely matches real-world scenarios, where deepfake detectors are required to identify deepfake sound generated by previously unseen generators.
Following this, we introduce the first track of the ESDD challenge, termed \textbf{ESDD in Unseen Generators}, which aims to develop robust detection models to unseen sound generators.

In the first track, although the sound generators in test sets are unseen, the underlying audio generation paradigms (i.e., TTA and ATA) have been encountered in the training process. 
In the second track, we consider a more challenging evaluation scenario, where the generation method is neither TTA nor ATA. Participants do not have specific information regarding the generation method used in the test data, it could be any approach except for TTA and ATA, such as multimodal audio generation \cite{qwen2.5-o,gpt-4o} and prompt-based audio editing \cite{audit}.
We refer to this evaluation setup as a ``black-box test'', which closely reflects real-world applications where detection systems must perform effectively without prior knowledge of generation paradigms.

Furthermore, considering that black-box data is typically difficult to obtain in practice, we simulate a low-resource setting by using a limited amount of such data for training, accounting for only 1\% of the total development set. 
We define the second track as \textbf{Black-Box Low-Resource ESDD}, aiming to evaluate the capability of models to generalize under extreme data scarcity while facing entirely unknown generation methods. 
To ensure the integrity and fairness, we will not disclose the generation paradigms used for evaluation in track 2 before the challenge concludes.

\subsection{Baseline Systems}
Following EnvSDD, we develop two systems as the baselines, namely AASIST and BEATs+AASIST, respectively.
AASIST \cite{aasist} is an end-to-end system that uses a novel heterogeneou stacking graph attention machanism \cite{heterogeneous} to learn acoustic features, which has been applied in various speech and singing voice deepfake detection challenges \cite{zhang2024svdd,asvspoof_2021}.
BEATs \cite{beats}, an audio foundation model, which was pre-trained on a large-scale environmental sound dataset (i.e., AudioSet-2M \cite{audioset}) with self-supervised learning, achieving superior performance on different downstream audio tasks.
Building on AASIST, BEATs+AASIST incorporates BEATs as the front-end, to extract high-level acoustic representations, achieving better performance than AASIST on EnvSDD.

For both systems, we use a batch size of 16 and the Adam optimizer with a weight decay of 0.0001. When training AASIST from scratch, the initial learning rate is set to 0.001. For fine-tuning the BEATs+AASIST model, the learning rate is reduced to 0.00001 to mitigate overfitting. The training process is limited to a maximum of 50 epochs, with early stopping applied if the validation loss does not improve for five consecutive epochs.

\subsection{Evaluation Metric}
We employ the equal error rate (EER) \cite{zhang2024svdd} as the metric to evaluate the performance of ESDD. 
Participants are expected to submit a score file in TXT format, containing confidence scores for each segmented sound clip. 
These scores indicate the system's confidence in determining whether a given sound clip originates from real-life.
In practical applications, users may set a decision threshold to convert these scores into binary outputs.
As the threshold increases, the false acceptance rate decreases, but the false rejection rate increases.
The EER is achieved when these two values are equal, providing a reliable and threshold-independent measure of system performance. A lower EER indicates better detection performance.

\begin{figure}
    \centering
    \includegraphics[width=1.05\linewidth]{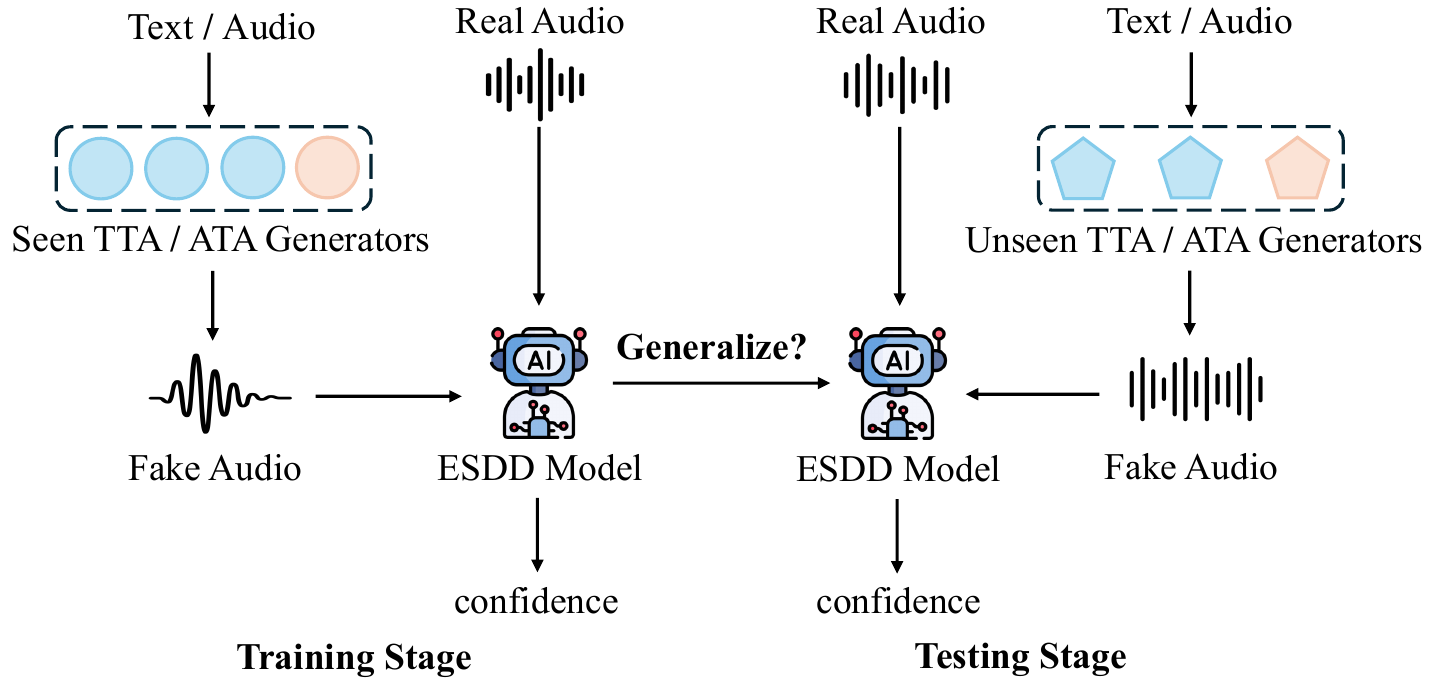}
    \caption{Overview of track 1: ESDD in Unseen Generators.}
    \label{fig:track1}
\end{figure}

\section{Track 1: ESDD in Unseen Generators}
\subsection{Objectives}
As shown in Fig.~\ref{fig:track1}, this track aims to encourage participants to develop robust ESDD systems for unseen TTA and ATA generators. The primary objective is to promote the creation of detection models that can effectively identify synthetic environmental sounds generated by unseen TTA and ATA systems. 
By focusing on the unseen-generator scenario, this track aims to simulate practical use cases where the source and generation models of deepfake audio are not seen during training. We seek to advance research in generalized deepfake detection and promote the design of models with strong cross-generator robustness and adaptability.

\subsection{Dataset}

For track 1, we directly use data from EnvSDD. Specifically, we first collect real samples from various open-source datasets, including UrbanSound8K \cite{urbansound}, TAU UAS 2019 Open Dev \cite{tut_uas}, TUT SED 2016 \cite{tutsed2016}, TUT SED 2017 \cite{tutsed2017_dev,tutsed2017_eva}, DCASE 2023 Task7 Dev \cite{foley} and Clotho \cite{clotho}, covering different real-life scenarios. 
All audio recordings are resampled to 16 kHz and split into 4s clips.
Then, as shown in Table~\ref{tab:tta and ata models}, we use 5 TTA models and 2 ATA models for generating the deepfake sound clips. 

\begin{table}
    \centering
    \caption{TTA and ATA models included in track 1.}
    \vspace{1mm}\renewcommand\arraystretch{0.9}{
    \setlength{\tabcolsep}{6.5mm}{
    \begin{tabular}{c|ccc}
        \toprule
        \# System & Type & Model \\
        \midrule
        G01 & TTA & AudioLDM \cite{audioldm}\\
        G02 & TTA & AudioLDM 2 \cite{audioldm2}\\
        G03 & TTA & AudioGen \cite{audiogen}\\
        G04 & ATA & AudioLDM \cite{audioldm}\\
        G05 & TTA & AudioLCM \cite{audiolcm}\\
        G06 & TTA & TangoFlux \cite{tangoflux}\\
        G07 & ATA & AudioLDM 2 \cite{audioldm2}\\
        \bottomrule
    \end{tabular}
    }}
    \label{tab:tta and ata models}
    \vspace{-5pt}
\end{table}

\begin{table}
    \centering
    \caption{Statistics of the dataset in track 1.}
    \vspace{1mm}\renewcommand\arraystretch{0.9}{
    \setlength{\tabcolsep}{2.5mm}{
    \begin{tabular}{c|cccc}
        \toprule
        \multirow{2}{*}{\textbf{Usage}} & \multicolumn{2}{c}{\textbf{Real}} & \multicolumn{2}{c}{\textbf{Fake}}\\
        & Sources & \# Clips & Methods & \# Clips \\
        \midrule
        Training & \cite{urbansound,tut_uas,tutsed2016,tutsed2017_dev,tutsed2017_eva} & 27,811 & G01$\sim$G04 & 111,244\\
        Validation & \cite{urbansound,tut_uas,tutsed2016,tutsed2017_dev,tutsed2017_eva} & 7,942 & G01$\sim$G04 & 31,768\\
        Evaluation & \cite{foley,clotho} & 478 & G05$\sim$G07 & 1,522\\
        Test & \cite{foley,clotho} & 1,000 & G05$\sim$G07 & 3,000  \\
        \bottomrule
    \end{tabular}
    }}
    \label{tab:statistics of track 1}
    \vspace{-5pt}
\end{table}

\begin{table}
    \vspace{-10pt}
    \centering
    \caption{The performance of baseline systems on track 1.}
    \vspace{1mm}\renewcommand\arraystretch{0.9}{
    \setlength{\tabcolsep}{0.8mm}{
    \begin{tabular}{c|c|ccc}
        \toprule
        \multirow{2}{*}{System} & \multirow{2}{*}{Params (M)} & \multicolumn{3}{c}{EER (\%)}\\
        & & Validation & Evaluation & Test \\
        \midrule
        AASIST & 0.30 & 0.92 & 15.26&	15.02 \\
        BEATs+AASIST & 90.73 & \textbf{0.10} & \textbf{14.21} &	\textbf{13.20}\\
        \bottomrule
    \end{tabular}
    }}
    \label{tab:baseline of track 1}
\end{table}

Table~\ref{tab:statistics of track 1} shows the specific statistics of the dataset for track 1. For training and validation splits, we use G01 to G04 for generating the deepfake data, while G05 to G07 are used for evaluation and test.
The evaluation set is intended for participants to select their models during the progress phase, while the test set is used for the final ranking. 
Specifically, the evaluation set is a randomly sampled subset of the test set, which ensures that the performance measured on the evaluation set remains aligned with the test distribution, while still preserving the integrity of the final assessment. 
All data and model checkpoints will be publicly available on Zenodo.

\subsection{Baseline Results}
Table~\ref{tab:baseline of track 1} presents the performance of AASIST and BEATs+AASIST on track 1.
For both systems, EERs on the evaluation and test sets are significantly higher than that on the validation set, highlighting the substantial challenge posed by unseen generators.
By incorporating the pretrained audio foundation model, BEATs, the EER on the evaluation set and test set is reduced by 1.05\% and 1.82\%, respectively. This trend consistent with our previous findings \cite{envsdd}.

\section{Track 2: Black-Box Low-Resource ESDD}
\subsection{Objectives}
\begin{figure}
    \centering
    \includegraphics[width=1\linewidth]{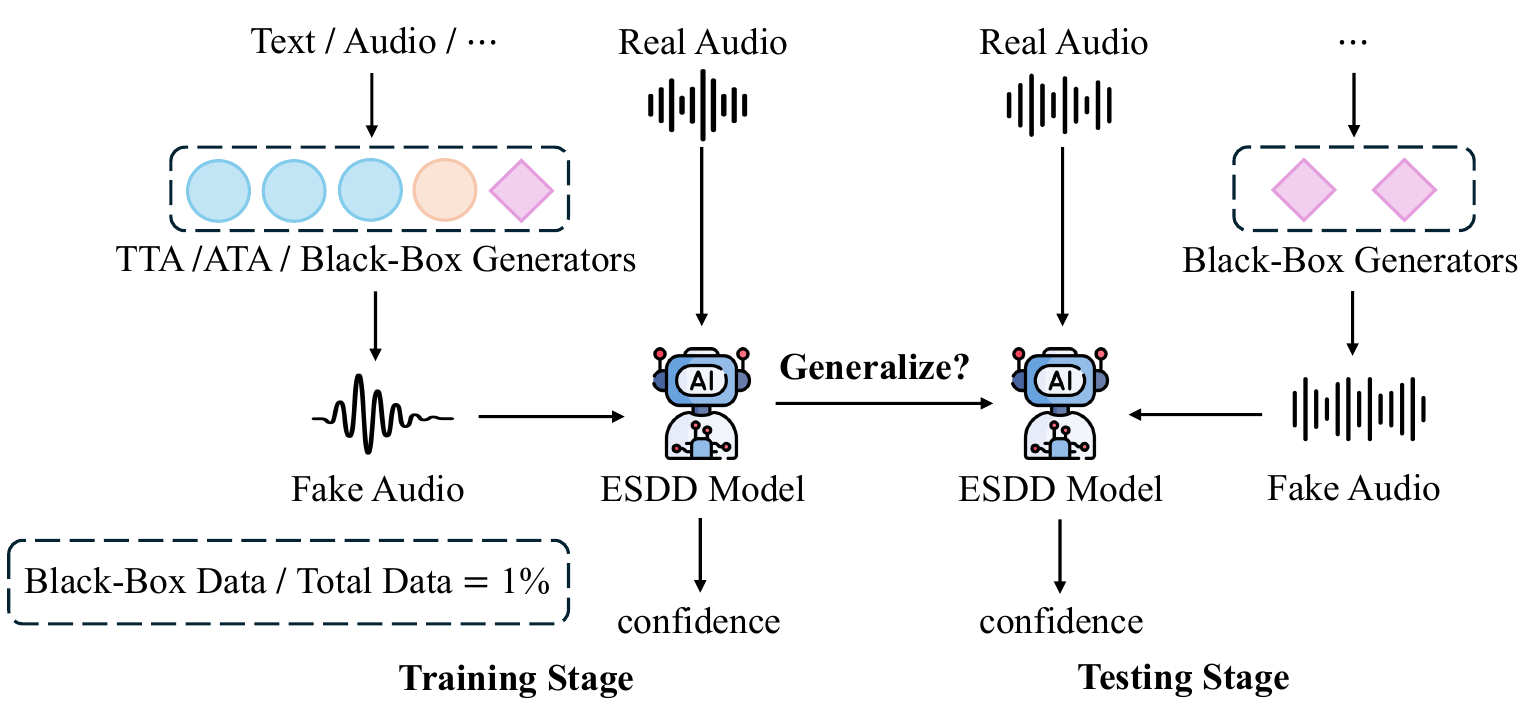}
    \vspace{-4mm}
    \caption{Overview of track 2: Black-Box Low-Resource ESDD.}
    \label{fig:track2}
    \vspace{-2mm}
\end{figure}

Fig.~\ref{fig:track2} illustrates the overview of track 2. 
The term ``black-box'' refers to the condition where participants have no prior knowledge of the specific generation methods used in testing, which may include any generative paradigms beyond TTA and ATA. The ``low-resource'' setting indicates that the amount of available black-box training data is severely limited, constituting only 1\% of the total training data. Together, this track presents a realistic and challenging scenario that simulates practical deepfake detection under extreme uncertainty and data scarcity.

\subsection{Dataset}
Table~\ref{tab:statistics of track 2} shows the statistics of the black-box data.
The scale of the black-box data is significantly smaller compared to that of track 1. 
We combine the training and validation sets from track 1 with the black-box data, where the latter accounts for only 1\% of the combined data. This limited proportion reflects the low-resource nature of the black-box setting and emphasizes the challenge of learning effective detection models under data constraints.

\subsection{Baseline Results}
We compare the performance of AASIST and BEATs+AASIST on track 2 in Table~\ref{tab:baseline of track 2}.
The results follow a similar trend to those observed in track 1, where the EER on the black-box test data is notably higher compared to the validation set. In should be noted that 90\% of the validation set consists of samples generated by seen TTA and ATA models, while the black-box test set contains entirely unseen generation paradigms. 
This performance gap highlights the difficulty of generalizing to unseen generation paradigms and underscores the importance of developing models that can effectively adapt to novel and unpredictable deepfake generation methods.

\section{Rules for System Development}
To ensure fairness in the competition, we have established the following rules for system development.
\begin{itemize}
    \item Participants are not allowed to use the evaluation set or test set for training purposes.
    \item The use of publicly available pre-trained models is allowed. However, the used models need to be clearly stated in the team's technical report. 
    \item Participants are encouraged to disclose the model parameters and training devices in their technical reports.
    \item Data augmentation is allowed. For example, open-source audio generation models can be used to synthesize fake data for training. However, the use of any internal generation models for any purposes is strictly not allowed. 
    \item The use of generators considered in evaluation and test sets (i.e., G05$\sim$G07) for training purposes is strictly not allowed.
    \item Participants must ensure that all training data originates from publicly available and ethically sourced datasets. Any use of proprietary, confidential, or unreleased data not explicitly permitted by the challenge organizers is strictly prohibited.
\end{itemize}

\begin{table}[t!]
    \centering
    \caption{Statistics of the black-box data in track 2.}
    \vspace{1mm}\renewcommand\arraystretch{0.8}{
    \setlength{\tabcolsep}{5mm}{
    \begin{tabular}{c|cc}
        \toprule
        \textbf{Usage} & \textbf{\# Real Clips} & \textbf{\# Fake Clips}\\
        \midrule
        Training & 270 & 1,083\\
        Validation & 90 & 361 \\
        Evaluation & 973 & 4,014\\
        Test & 1,994 & 7,980 \\
        \bottomrule
    \end{tabular}
    }}
    \label{tab:statistics of track 2}
     \vspace{-2mm}
\end{table}

\begin{table} [t!]
    \centering
    \caption{The performance of baseline systems on track 2.}
    \vspace{1mm}\renewcommand\arraystretch{0.8}{
    \setlength{\tabcolsep}{0.8mm}{
    \begin{tabular}{c|c|ccc}
        \toprule
        \multirow{2}{*}{System} & \multirow{2}{*}{Params (M)} & \multicolumn{3}{c}{EER (\%)}\\
        & & Validation & Evaluation & Test \\
        \midrule
        AASIST & 0.30 & 1.11 & 15.72 &	15.40 \\
        BEATs+AASIST & 90.73 & \textbf{0.16} & \textbf{12.64} & \textbf{12.48}\\
        \bottomrule
    \end{tabular}
    }}
    \label{tab:baseline of track 2}
    \vspace{-2mm}
\end{table}

\section{Registration Process}
The following Google Form will be used by participating teams for registration of their respective teams in the challenge\footnote{Registration Link: https://forms.gle/hvhPs5Ru7skUXY7a6}.

\section{Submission of Results}
We require participants to submit their test set predictions and system descriptions, which will be published on the challenge website. To promote reproducibility, we strongly encourage participants to open-source both their training and inference code.

We use CodaBench \cite{codabench} for results submission.  
Specifically, we will open two CodaBench links for track 1\footnote{Codabench for Track 1: https://www.codabench.org/competitions/10014/} and track 2\footnote{Codabench for Track 2: https://www.codabench.org/competitions/10015/}, respectively.
Participants are expected to compute and submit text files including the ID and confidence scores in the following format:

\begin{itemize}
    \item \textit{A8FzlYLvB1pteancdKWk.wav, -0.0524}
    \item \textit{...}
    \item \textit{BONmE6pXle1kc2Rf8ULJ.wav, -5.1601}
\end{itemize}

It should be noted that during ranking phase, each team is allowed to submit at most three results per track. This limit is imposed to ensure fairness and prevent excessive tuning based on test set feedback. After the competition concludes, we will release the metadata of the test sets to promote reproducibility, and future research in environmental sound deepfake detection. 
In addition, during ranking phase, the participants are required to submit configuration file(s) in an expected template to the challenge organizers (one configuration file for one submitted system). The template of the configuration file will be shared in the challenge Google group\footnote{Google group: https://groups.google.com/g/esdd-challenge-2026}.


The ESDD 2026 challenge is one of the Grand Challenges in ICASSP 2026. The top ranked participants of the challenge will be invited to submit challenge papers to ICASSP.

\section{Important Dates}
Tentative timeline of the challenge following ICASSP 2026 Grand Challenge submission:
\begin{itemize}
\item Launch of the challenge: September 01, 2025
\item Registration deadline: October 15, 2025
\item Progress phase: September 01, 2025 ~ November 16, 2025
\item Test sets release: November 17, 2025
\item Ranking phase: November 17, 2025 ~ November 26, 2025
\item Technical report submission: 26 November 2025
\item Leaderboard release: November 27, 2025
\item Two-page paper due (by invitation only): December 07, 2025
\item Two-page paper acceptance notification: January 11, 2026
\item Camera-ready 2-page papers due: January 18, 2026
\end{itemize}
\balance
\bibliographystyle{IEEEbib}
\bibliography{refs}

@inproceedings{foley,
  title={{Foley sound synthesis at the DCASE 2023 challenge}},
  author={Choi, Keunwoo and Im, Jaekwon and Heller, Laurie and Mcfee, Brian and Imoto, Keisuke and Okamoto, Yuki and Lagrange, Mathieu and Takamichi, Shinnosuke},
  booktitle={Workshop on Detection and Classification of Acoustic Scenes and Events},
  year={2023}
}

@inproceedings{add_challenge,
  title={{ADD} 2023: {The} Second Audio Deepfake Detection Challenge},
  author={Yi, Jiangyan and Tao, Jianhua and Fu, Ruibo and Yan, Xinrui and Wang, Chenglong and Wang, Tao and Zhang, Chu Yuan and Zhang, Xiaohui and Zhao, Yan and Ren, Yong and others},
  booktitle={CEUR Workshop Proceedings},
  volume={3597},
  pages={125--130},
  year={2023}
}

@article{deepfake-review-2,
  title={{Audio deepfake detection: A survey}},
  author={Yi, Jiangyan and Wang, Chenglong and Tao, Jianhua and Zhang, Xiaohui and Zhang, Chu Yuan and Zhao, Yan},
  journal={CoRR},
  year={2023}
}

@article{deepfake-review-1,
  title={The effect of deep learning methods on deepfake audio detection for digital investigation},
  author={Mcuba, Mvelo and Singh, Avinash and Ikuesan, Richard Adeyemi and Venter, Hein},
  journal={Procedia Computer Science},
  volume={219},
  pages={211--219},
  year={2023},
  publisher={Elsevier}
}

@inproceedings{envsdd,
  title={{EnvSDD}: Benchmarking environmental sound deepfake detection},
  author={Yin, Han and Xiao, Yang and Das, Rohan Kumar and Bai, Jisheng and Liu, Haohe and Wang, Wenwu and Plumbley, Mark D},
  booktitle={Proc. Interspeech},
  year={2025}
}

@article{asvspoof_2021,
  title={Asvspoof 2021: Towards spoofed and deepfake speech detection in the wild},
  author={Liu, Xuechen and Wang, Xin and Sahidullah, Md and Patino, Jose and Delgado, H{\'e}ctor and Kinnunen, Tomi and Todisco, Massimiliano and Yamagishi, Junichi and Evans, Nicholas and Nautsch, Andreas and others},
  journal={IEEE/ACM Transactions on Audio, Speech, and Language Processing},
  volume={31},
  pages={2507--2522},
  year={2023},
  publisher={IEEE}
}

@inproceedings{asvspoof5,
  title={ASVspoof 5 Challenge: Advanced ResNet Architectures for Robust Voice Spoofing Detection},
  author={Dao, Anh-Tuan and Rouvier, Mickael and Matrouf, Driss},
  booktitle={Proc. ASVspoof},
  pages={163--169},
  year={2024}
}

@article{asvspoof2017,
  title={ASVspoof: The automatic speaker verification spoofing and countermeasures challenge},
  author={Wu, Zhizheng and Yamagishi, Junichi and Kinnunen, Tomi and Hanil{\c{c}}i, Cemal and Sahidullah, Mohammed and Sizov, Aleksandr and Evans, Nicholas and Todisco, Massimiliano and Delgado, Hector},
  journal={IEEE Journal of Selected Topics in Signal Processing},
  volume={11},
  number={4},
  pages={588--604},
  year={2017},
  publisher={IEEE}
}

@inproceedings{beats,
  title={{BEATs}: Audio pre-training with acoustic tokenizers},
  author={Chen, Sanyuan and Wu, Yu and Wang, Chengyi and Liu, Shujie and Tompkins, Daniel and Chen, Zhuo and Che, Wanxiang and Yu, Xiangzhan and Wei, Furu},
  booktitle={Proc. International Conference on Machine Learning (ICML)},
  pages={5178--5193},
  year={2023}
}

@article{qwen2.5-o,
  title={Qwen2.5-omni technical report},
  author={Xu, Jin and Guo, Zhifang and He, Jinzheng and Hu, Hangrui and He, Ting and Bai, Shuai and Chen, Keqin and Wang, Jialin and Fan, Yang and Dang, Kai and others},
  journal={arXiv preprint:2503.20215},
  year={2025}
}

@article{gpt-4o,
  title={Gpt-4o system card},
  author={Hurst, Aaron and Lerer, Adam and Goucher, Adam P and Perelman, Adam and Ramesh, Aditya and Clark, Aidan and Ostrow, AJ and Welihinda, Akila and Hayes, Alan and Radford, Alec and others},
  journal={arXiv preprint:2410.21276},
  year={2024}
}

@inproceedings{audit,
  title={Audit: Audio editing by following instructions with latent diffusion models},
  author={Wang, Yuancheng and Ju, Zeqian and Tan, Xu and He, Lei and Wu, Zhizheng and Bian, Jiang and others},
  booktitle={Advances in Neural Information Processing Systems (NIPS)},
  volume={36},
  year={2023},
  pages={71340--71357},
  
}

@inproceedings{aasist,
  title={{AASIST}: Audio anti-spoofing using integrated spectro-temporal graph attention networks},
  author={Jung, Jee-weon and Heo, Hee-Soo and Tak, Hemlata and Shim, Hye-jin and Chung, Joon Son and Lee, Bong-Jin and Yu, Ha-Jin and Evans, Nicholas},
  booktitle={Proc. International Conference on Acoustics, Speech and Signal Processing (ICASSP)},
  pages={6367--6371},
  year={2022},
  organization={IEEE}
}

@inproceedings{heterogeneous,
  title={Heterogeneous graph attention network},
  author={Wang, Xiao and Ji, Houye and Shi, Chuan and Wang, Bai and Ye, Yanfang and Cui, Peng and Yu, Philip S},
  booktitle={The World Wide Web Conference},
  pages={2022--2032},
  year={2019}
}

@inproceedings{audioset,
  title={Audio set: An ontology and human-labeled dataset for audio events},
  author={Gemmeke, Jort F and Ellis, Daniel PW and Freedman, Dylan and Jansen, Aren and Lawrence, Wade and Moore, R Channing and Plakal, Manoj and Ritter, Marvin},
  booktitle={International Conference on Acoustics, Speech and Signal Processing (ICASSP)},
  pages={776--780},
  year={2017},
  organization={IEEE}
}

@inproceedings{urbansound,
  title={A dataset and taxonomy for urban sound research},
  author={Salamon, Justin and Jacoby, Christopher and Bello, Juan Pablo},
  booktitle={Proc. ACM International Conference on Multimedia (ACM MM)},
  pages={1041--1044},
  year={2014}
}

@article{tut_uas,
  title={{TAU urban acoustic scenes 2019 openset, development dataset}},
  author={Heittola, Toni and Mesaros, A and Virtanen, T},
  year={2019},
  journal={Zenodo},
  note={doi:{\color{blue}\href{https://doi.org/10.5281/zenodo.2591503}{https://doi.org/10.5281/zenodo.2591503}}}
}

@inproceedings{tutsed2016,
  title={{TUT database for acoustic scene classification and sound event detection}},
  author={Mesaros, Annamaria and Heittola, Toni and Virtanen, Tuomas},
  booktitle={Proc. European Signal Processing Conference (EUSIPCO)},
  pages={1128--1132},
  year={2016},
}

@article{tutsed2017_dev,
  title={{TUT sound events 2017, development dataset}},
  author={Mesaros, Annamaria1 and Heittola, Toni1 and Virtanen, Tuomas},
  year={2017},
  journal={Zenodo},
note={doi:{\color{blue}\href{https://doi.org/10.5281/zenodo.814831}{https://doi.org/10.5281/zenodo.814831}}}
}

@article{tutsed2017_eva,
  title={{TUT sound events 2017, evaluation dataset}},
  author={Mesaros, Annamaria1 and Heittola, Toni1 and Virtanen, Tuomas},
  year={2017},
  journal={Zenodo},
note={doi:{\color{blue}\href{https://doi.org/10.5281/zenodo.1040179}{https://doi.org/10.5281/zenodo.1040179}}}
}

@inproceedings{clotho,
  title={{Clotho: an audio captioning dataset}},
  author={Drossos, Konstantinos and Lipping, Samuel and Virtanen, Tuomas},
  booktitle={Proc. International Conference on Acoustics, Speech and Signal Processing (ICASSP)},
  pages={736--740},
  year={2020},
}

@inproceedings{audioldm,
  title={{AudioLDM: Text-to-audio generation with latent diffusion models}},
  author={Liu, Haohe and Chen, Zehua and Yuan, Yi and Mei, Xinhao and Liu, Xubo and Mandic, Danilo and Wang, Wenwu and Plumbley, Mark D},
  booktitle={Proc. International Conference on Machine Learning (ICML)},
  pages={21450--21474},
  year={2023},
}

@article{audioldm2,
  title={{AudioLDM 2: Learning holistic audio generation with self-supervised pretraining}},
  author={Liu, Haohe and Yuan, Yi and Liu, Xubo and Mei, Xinhao and Kong, Qiuqiang and Tian, Qiao and Wang, Yuping and Wang, Wenwu and Wang, Yuxuan and Plumbley, Mark D},
  journal={IEEE/ACM Transactions on Audio, Speech, and Language Processing},
  year={2024},
  volume={32},
  pages={2871--2883},
  publisher={IEEE}
}

@inproceedings{audiogen,
  title={{AudioGen: Textually guided audio generation}},
  author={Kreuk, Felix and Synnaeve, Gabriel and Polyak, Adam and Singer, Uriel and D{\'e}fossez, Alexandre and Copet, Jade and Parikh, Devi and Taigman, Yaniv and Adi, Yossi},
  booktitle={Proc. International Conference on Learning Representations (ICLR)},
  pages={1--16},
  year={2023}
}

@inproceedings{audiolcm,
  title={{AudioLCM: Efficient and high-quality text-to-audio generation with minimal inference steps}},
  author={Liu, Huadai and Huang, Rongjie and Liu, Yang and Cao, Hengyuan and Wang, Jialei and Cheng, Xize and Zheng, Siqi and Zhao, Zhou},
  booktitle={Proc. ACM International Conference on Multimedia (ACM MM)},
  pages={7008--7017},
  year={2024}
}

@article{tangoflux,
  title={{TangoFlux: Super fast and faithful text to audio generation with flow matching and clap-ranked preference optimization}},
  author={Hung, Chia-Yu and Majumder, Navonil and Kong, Zhifeng and Mehrish, Ambuj and Valle, Rafael and Catanzaro, Bryan and Poria, Soujanya},
  journal={arXiv preprint: 2412.21037},
  year={2024}
}

@article{codabench,
  title={Codabench: Flexible, easy-to-use, and reproducible meta-benchmark platform},
  author={Xu, Zhen and Escalera, Sergio and Pav{\~a}o, Adrien and Richard, Magali and Tu, Wei-Wei and Yao, Quanming and Zhao, Huan and Guyon, Isabelle},
  journal={Patterns},
  volume={3},
  number={7},
  year={2022},
  publisher={Elsevier}
}

@article{zhang2024svdd,
  title={{SVDD Challenge 2024}: A Singing Voice Deepfake Detection Challenge Evaluation Plan},
  author={Zhang, You and Zang, Yongyi and Shi, Jiatong and Yamamoto, Ryuichi and Han, Jionghao and Tang, Yuxun and Toda, Tomoki and Duan, Zhiyao},
  journal={CoRR},
  year={2024}
}

@inproceedings{zang2024ctrsvdd,
  title={{CtrSVDD}: A Benchmark Dataset and Baseline Analysis for Controlled Singing Voice Deepfake Detection},
  author    = {Yongyi Zang and Jiatong Shi and You Zhang and Ryuichi Yamamoto and Jionghao Han and Yuxun Tang and Shengyuan Xu and Wenxiao Zhao and Jing Guo and Tomoki Toda and Zhiyao Duan},
  year      = {2024},
  booktitle = {Proc. Interspeech},
  pages     = {4783--4787},
  doi       = {10.21437/Interspeech.2024-2242},
  issn      = {2958-1796},
}
\end{document}